\documentclass[twocolumn, 5p]{elsarticle}

\usepackage{graphicx}
\usepackage{xcolor}
\usepackage{color}
\usepackage{amsmath,amsfonts,amssymb}
\usepackage{latexsym}
\usepackage[utf8]{inputenc}

\title{Emergence of dynamical dark energy from polynomial $f(R)$ theory in Palatini formalism}

\author{Marek Szyd{\l}owski}
\ead{marek.szydlowski@uj.edu.pl}
\address{Astronomical Observatory, Jagiellonian University, Orla 171, 30-244 Krakow, Poland}
\address{Mark Kac Complex Systems Research Centre, Jagiellonian University, {\L}ojasiewicza 11, 30-348 Krak{\'o}w, Poland}
\author{Aleksander Stachowski}
\ead{aleksander.stachowski@doctoral.uj.edu.pl}
\address{Astronomical Observatory, Jagiellonian University, Orla 171, 30-244 Krakow, Poland}
\author{Andrzej Borowiec}
\ead{andrzej.borowiec@ift.uni.wroc.pl}
\address{Institute for Theoretical Physics, Wroclaw University, pl. Maxa Borna 9, 50-204 Wroclaw, Poland}

\begin{document}

\begin{abstract}
We consider FRW cosmology in $f(R)= R+ \gamma R^2+\delta R^3$ modified  framework. The Palatini approach reduces its dynamics to the simple generalization of Friedmann equation. Thus we study the dynamics in two-dimensional phase space with some details. After reformulation of the model in the Einstein frame, it reduces to the FRW cosmological model with a homogeneous scalar field and vanishing kinetic energy term. This potential determines the running cosmological constant term as a function of the Ricci scalar. As a result we obtain the emergent dark energy parametrization from the covariant theory. We study also singularities of the model and demonstrate that in the Einstein frame some undesirable singularities disappear.
\end{abstract}

\maketitle

\section{Introduction}

A variety of explanations have been proposed for the Universe accelerating expansion at the current epoch. Among them, the idea of positive cosmological constant $\Lambda$, as the simplest candidates, seems to be viable. However, it is only an economical description (with the help of one free parameter)  of observational facts rather than  an effective explanation.
The simplest alternative candidate for the constant cosmological parameter being a key element in the standard cosmological model (called $\Lambda$CDM model) is a time-dependent (or running) cosmological term. It is crucial for avoiding fine-tuning and coincidence problems \cite{Martin:2012bt,Peebles:2002gy}.

It would be nice to derive the dynamics of the running cosmological term as an emergent phenomenon from a more fundamental theory, for example from the string theory or from the first principles of quantum mechanics \cite{Stachowski:2016zpq}.
In this context, it is important to formulate a dynamical cosmological term without violating the covariance of the action. For example, models with a slowly rolling homogeneous cosmological scalar field, provide a popular alternative to the standard time-independent cosmological constant. We can study the simultaneous evolution of the background expansion and an evolution of the scalar field  with the self-interacting potential \cite{Ratra:1987rm}.

 In this paper we are going to push forward an idea of the emergent running cosmological term from a covariant theory \cite{Stachowski:2016dfi}. Parametrization of the cosmological term is derived directly from a formulation of the model in the Einstein frame by means of the Palatini variational approach. In analogy with Starobinsky's purely metric formulation  \cite{Starobinsky:1980te}, we obtain the parametrization of the cosmological term directly from the potential of the scalar field which appears after formulation of the specific FRW  model in the Einstein frame. As a next step we investigate the dynamics of the model with such a form of the dark energy.

In this letter, we demonstrate how $f(R)$ model is modified in the Palatini formulation. Our construction provides a simple model of an evolving dark energy (running cosmological term) to explain a dynamical relaxation of the vacuum energy (gravitational repulsive pressure) to a very small value today (cosmological constant problem \cite{Steinhardt:2006bf}). This model, when studied in the Einstein frame, leads also to a small deviation from the $w=-1$ prediction of the non-running dark energy.

\section{Cosmological equations for the polynomial $f(R)$ theory in the Palatini formalism}

The Palatini gravity action for $f(\hat{R})$ gravity is given by
\begin{equation}
S=S_{\text{g}}+S_{\text{m}}=\frac{1}{2}\int \sqrt{-g}f(\hat{R}) d^4 x+S_{\text{m}},\label{action}
\end{equation}
where $\hat{R}$ is the generalized Ricci scalar \cite{Allemandi:2004wn,Olmo:2011uz}. From the action (\ref{action}) we get 
\begin{equation}
f'(\hat{R})\hat{R}_{(\mu\nu)}-\frac{1}{2}f(\hat{R})g_{\mu\nu}=T_{\mu\nu},\label{structural}
\end{equation}
\begin{equation}
\hat{\nabla}_\alpha(\sqrt{-g}f'(\hat{R})g^{\mu\nu})=0,\label{con}
\end{equation}
where $T_{\mu\nu}$ is energy momentum tensor and $\hat{\nabla}_\alpha$ is the covariant derivative calculated with respect to $\Gamma$.

If we take the trace of Eq. (\ref{structural}), we get a structural equation, which is given by
\begin{equation}
f'(\hat{R})\hat{R}-2 f(\hat{R})=T.\label{structural2}
\end{equation}
where $T=g^{\mu\nu}T_{\mu\nu}$.
We assume the FRW metric in the following form
\begin{equation}\label{frw}
ds^2=dt^2-a^2(t)\left[\frac{1}{1-kr^2}dr^2+r^2(d\theta^2+\sin^2\theta d\phi^2)\right],
\end{equation}
where $a(t)$ is the scale factor, $k$ is a constant of spatial curvature ($k=0, \pm 1$) and $t$ is the cosmological time. Thereafter, we assume the flat model ($k=0$).

We assume the energy-momentum tensor for perfect fluid
\begin{equation}
T^\mu_\nu=\text{diag}(-\rho,p,p,p),
\end{equation} 
where $p=w\rho$ with $w=const$. The conservation condition $T_{\nu;\mu}^{\mu}=0$ \cite{Koivisto:2005yk} gives 
\begin{equation}
\dot\rho_\text{m}=-3(1+w)H\rho_\text{m},\label{conservation}
\end{equation}
where $H$ is the Hubble function and $\rho_\text{m}$ is the density of baryonic and dark matter which is assumed to be in the form of dust ($w=0$).

In our paper the function $f(\hat{R})$ is assumed in the polynomial form as
\begin{equation}\label{lag}
f(\hat{R})=\sum_{i=1}^n \gamma_i \hat{R}^i,
\end{equation}
where $\gamma_i$ are some dimensionful parameters.

Therefore, we introduce more convenient dimensionless functions and parameters
\begin{multline}
\Omega_\text{R}=\frac{\hat{R}}{3H_0^2},\qquad \Omega_{\gamma_i} =3^{i-1}\gamma_i H_0^{2(i-1)}, \\ 
\Omega_{\text{tot}} =\Omega_{\text{m},0}a^{-3}+\Omega_{\Lambda,0}, \qquad
b =f'(\hat{R})=\sum_{i=1}^n i\Omega_{\gamma_i} \Omega_R^{i-1},\;\; \\d=-3\left(\sum_{i=1}^n (i-2)\Omega_{\gamma_i} \Omega_R^{i-1}+\frac{4\Omega_\Lambda}{\Omega_R}\right)\\
\times \frac{\sum_{i=1}^n i(i-1)\Omega_{\gamma_i} \Omega_R^{i-1}}{\sum_{i=1}^n i(i-2)\Omega_{\gamma_i} \Omega_R^{i-1}}.
\end{multline}
where $H_0$ is the present value of Hubble function, $\Omega_\text{m,0}=\frac{\rho_\text{\text{m},0}}{3H_0^2}$,  
$\Omega_{\Lambda,0}=\frac{\rho_{\Lambda,0}}{3H_0^2}$.\footnote{For the sake of generality (following the standard cosmological model) the presence of the cosmological constant is also assumed.}

For the function (\ref{lag}) the structural equation (\ref{structural2}) is in the following form
\begin{equation}\label{sol}
\sum_{i=1}^n (i-2)\Omega_{\gamma_i} \Omega_R^{i}=-\Omega_\text{m}-4\Omega_\Lambda.
\end{equation}

The Friedmann equation for the function (\ref{lag}) has the following form
\begin{multline}
\frac{H^2}{H_0^2}=\frac{b^2}{\left(b+\frac{d}{2}\right)^2}\\
\times \left[\frac{1}{2b}\left[\sum_{i=1}^n \Omega_{\gamma_i} \Omega_R^{i-1}\left(\Omega_R-2i\Omega_{\text{tot}}\right)+\Omega_{\text{tot}}-3\Omega_\Lambda\right]+\Omega_{\text{tot}}\right].
\label{friedmann2}
\end{multline}

\section{Singularities in the polynomial $f(R)$ theory in the Palatini formalism}

The Friedmann equation (\ref{friedmann2}) can be rewritten in an equivalent form 
\begin{equation}
{a'}^2=-2V(a),\label{friedmann}
\end{equation} 
where $'=\frac{d}{d\tau}=\frac{|b+d/2|}{|b|}\frac{d}{d t}$ is a new parametrization of time (this parametrization is not a diffeomorphism) and
\begin{multline}
V(a)=-\frac{H_0^2 a^2}{2}\\ 
\times \left[\frac{1}{2b}
\left[\sum_{i=1}^n \Omega_{\gamma_i} \Omega_R^{i-1}\left(\Omega_R-2i\Omega_{\text{tot}}\right)+\Omega_{\text{tot}}-3\Omega_\Lambda\right]+\Omega_{\text{tot}}\right].\label{potential}
\end{multline}

The potential $V(a)$ can be used to construction of a phase space portrait.
In this case the phase space is two-dimensional
\begin{equation}
	\left\{ (a,a') \colon \frac{a'^2}{2}+V(a)=-\frac{k}{2} \right\}.
\end{equation}
The dynamical system has the following form
\begin{align}
a' &= x,\label{dota} \\
x' &= -\frac{\partial V(a)}{\partial a} .\label{dotx} 
\end{align}
We assume that the potential function, except some isolated (singular) points, belongs to the class $C^2(\mathbb{R}_+)$.

\begin{figure}[ht]
\centering
\includegraphics[scale=0.5]{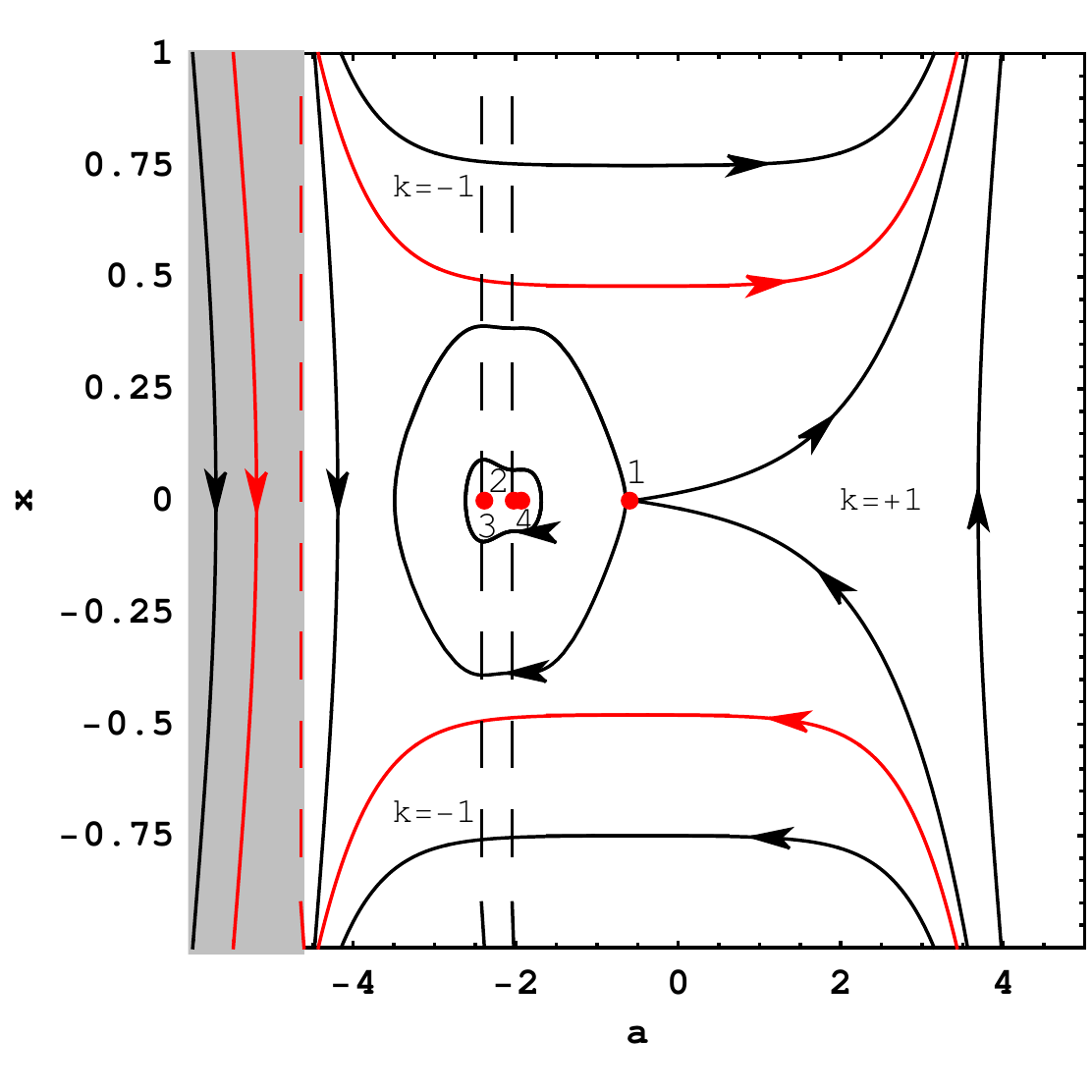}
\caption{The phase portrait for system (\ref{dota})-(\ref{dotx}) with $f(\hat{R})=\hat{R}+\gamma \hat{R}^2+\delta\hat{R}^3 $, where $\gamma=10^{-6}\left[\frac{\text{s}^2\text{Mpc}^2}{\text{km}^2}\right]$ and $\delta=-10^{-14}\left[\frac{\text{s}^4\text{Mpc}^4}{\text{km}^4}\right]$.  Critical points (1), (2), (3) and (4) represent the static Einstein universes. Critical points (1) and (2) are the saddle type and critical points (3) and (4) are the center type. The red dashed line presents the sudden singularity. The black dashed lines present the freeze singularities. The grey color marks the non-physical domain ($f'(R)<0$). The reds trajectories represent the path of evolution for the flat universe. These trajectories seperate the domain with the negative curvature ($k=-1$) from the domain with the positive curvature ($k=+1$). The scale factor is expressed in the logarithmic scale.}
\label{fig8}
\end{figure}

\begin{figure}[ht]
\centering
\includegraphics[scale=0.5]{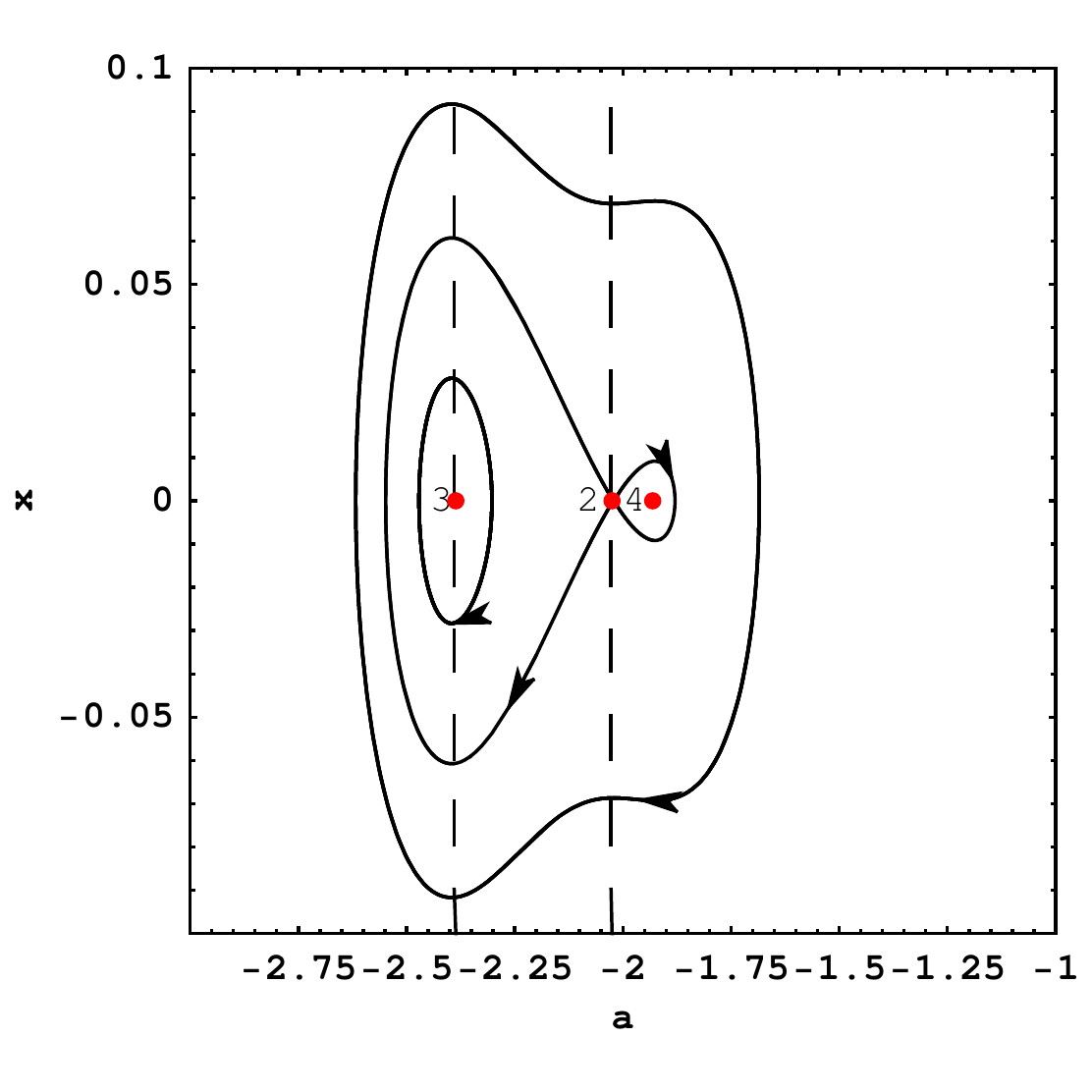}
\caption{The zoomed region of figure \ref{fig8}. The behaviour of trajectories in the neighbourhood of critical points (2), (3) and (4) which represent the static Einstein universes. Critical point (2) is the saddle type and critical points (3) and (4) are the center type. The black dashed lines present the freeze singularities. The scale factor is expressed in the logarithmic scale. The homoclinic orbits represent the bouncing models, which evolution starts and ends at the Einstein Universe (critical point 2). In the domain bounded by the homoclinic orbits the oscilating models present without the initial singularity.}
\label{fig9}
\end{figure}

\begin{figure}[ht]
\centering
\includegraphics[scale=0.5]{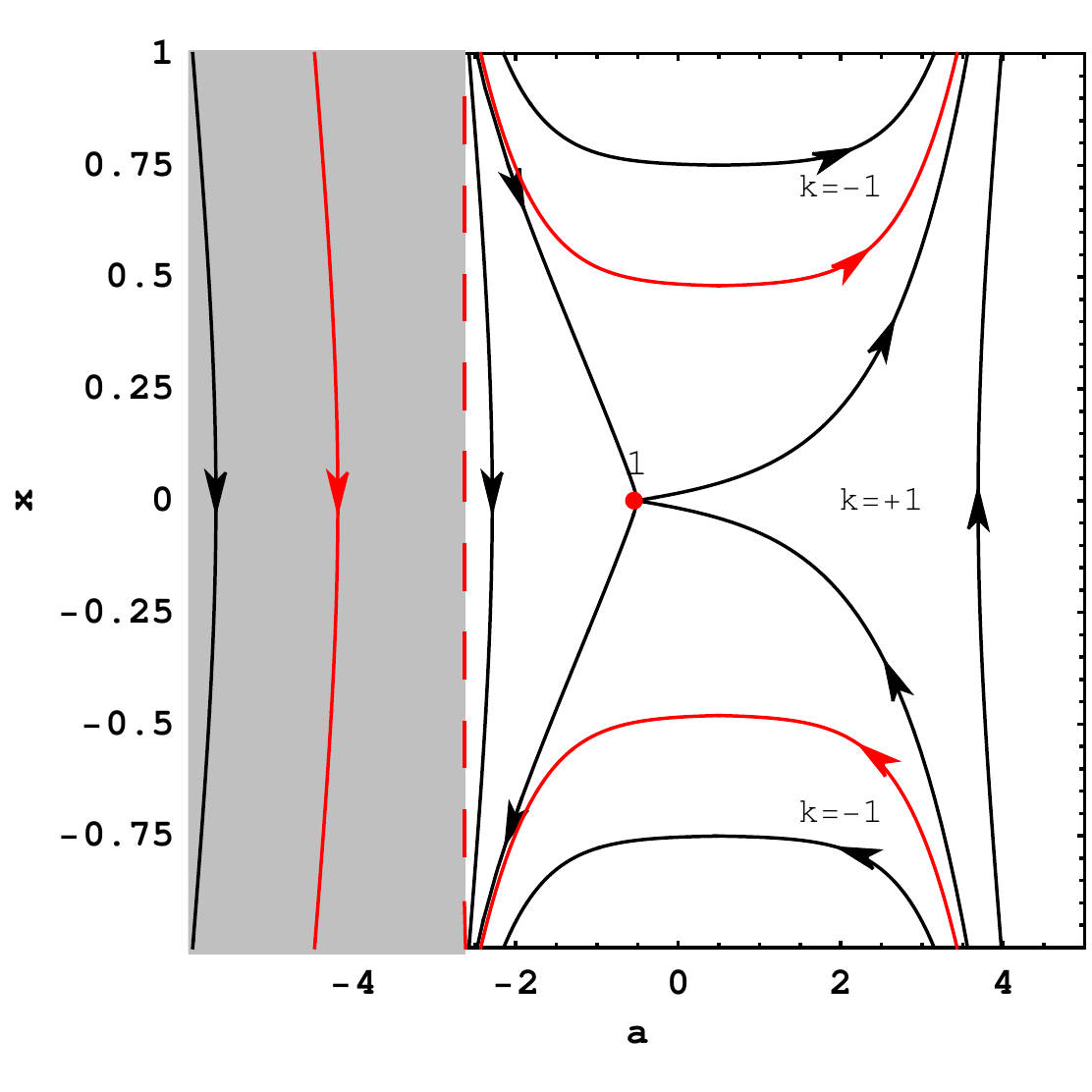}
\caption{The phase portrait for system (\ref{dota})-(\ref{dotx}) with $f(\hat{R})=\hat{R}+\gamma\hat R+\delta\hat{R}^3 $, where $\gamma=-10^{-6}\left[\frac{\text{s}^2\text{Mpc}^2}{\text{km}^2}\right]$ and $\delta=-10^{-14}\left[\frac{\text{s}^4\text{Mpc}^4}{\text{km}^4}\right]$.  Critical point (1), which is the saddle type, represents the static Einstein universe. The red dashed line presents the sudden singularity. The grey color presents the non-physical domain ($f'(R)<0$). The reds trajectories represent the path of evolution for the flat universe. These trajectories seperate the domain with the negative curvature ($k=-1$) from the domain with the positive curvature ($k=+1$). The scale factor is expressed in the logarithmic scale.}
\label{fig10}
\end{figure}

The example phase portraits for the dynamical system (\ref{dota})-(\ref{dotx}) are presented in figures \ref{fig8}, \ref{fig9} and \ref{fig10}.

The evolution of a universe can be treated as a motion of a fictitious particle of unit mass in the potential $V(a)$. Here $a(t)$ plays the role of a positional variable. Equation of motion (\ref{dotx}) assumes the form analogous to the Newtonian equation of motion. In this case the lines $\frac{x^2}{2}+V(a)=-\frac{k}{2}$ represent possible evolutions of the universe for different initial conditions.

In our model, there are two types of singularities: the freeze and sudden singularity. They are a consequence of the Palatini formalism. We get the freeze singularity when $b+d/2=0$. The sudden singularity appears when $b=0$ or $b+d/2$ is equal the infinity.

\begin{figure}[ht]
\centering
\includegraphics[scale=0.55]{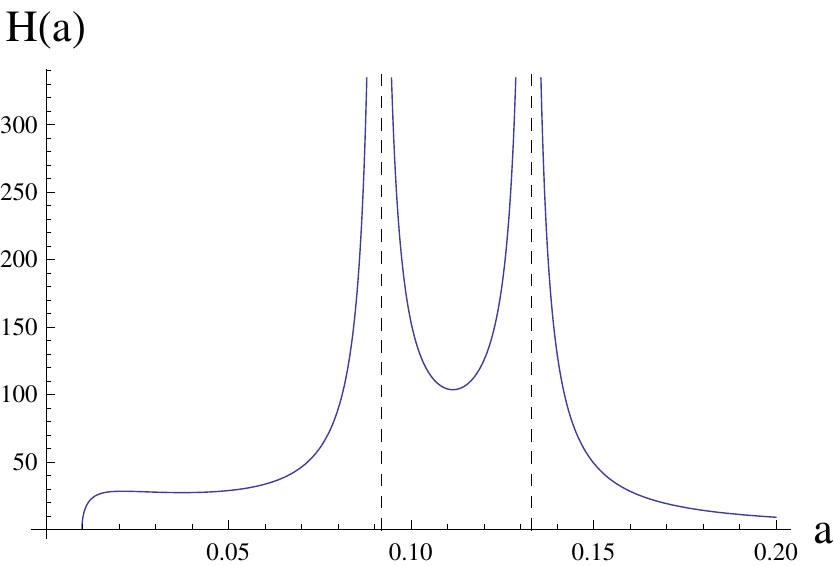}
\caption{The evolution of the Hubble function for $f(\hat{R})=\hat{R}+\gamma \hat{R}^2+\delta\hat{R}^3 $, where $\gamma=10^{-6}\left[\frac{\text{s}^2\text{Mpc}^2}{\text{km}^2}\right]$ and $\delta=-10^{-14}\left[\frac{\text{s}^4\text{Mpc}^4}{\text{km}^4}\right]$. The black dashed lines present the freeze singularities. Note that the singularity of the big bang type does not appear here. Units of $H(a)$ are expressed in $\frac{\text{100 km}}{\text{s Mpc}}$.}
\label{fig4}
\end{figure}

\begin{figure}[ht]
\centering
\includegraphics[scale=0.55]{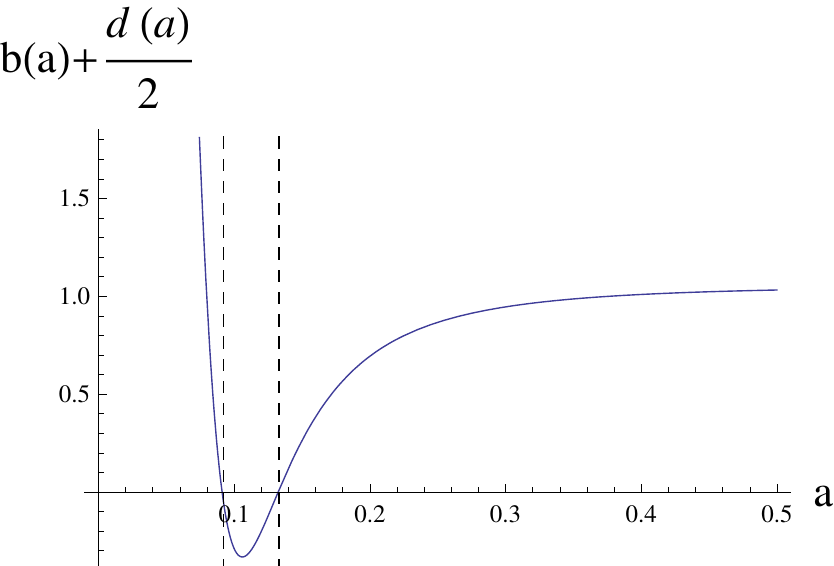}
\caption{The evolution of $b(a)+\frac{d(a)}{2}$. For values of scale factor, for which equation $b(a)+\frac{d(a)}{2}=0$ has roots, the freeze singularities appear (black dashed lines). This figure corresponds with figure \ref{fig4}.}
\label{fig5}
\end{figure}

\begin{figure}[ht]
\centering
\includegraphics[scale=0.55]{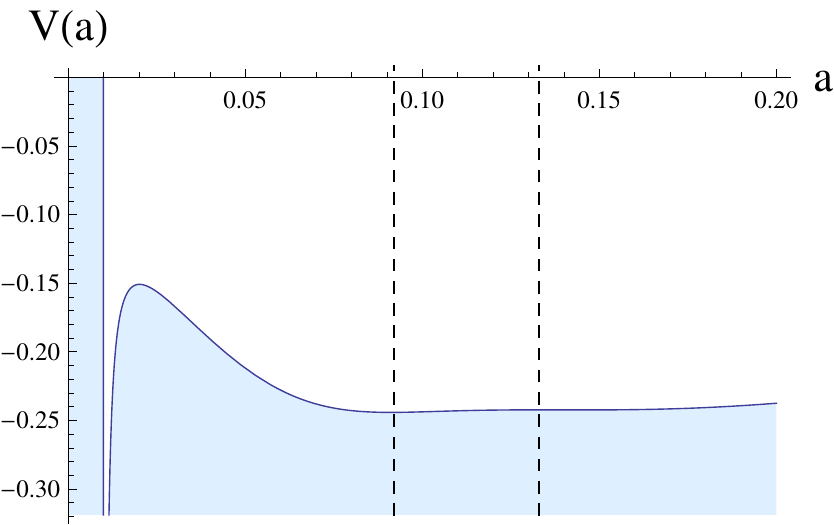}
\caption{The evolution of $V(a)$. This figure corresponds with figure \ref{fig4}. The black dashed lines present the freeze singularities. The potential is regular at these singularities while its higher derivatives blows up. The pole of $V(a)$ represents the sudden singularity.}
\label{fig6}
\end{figure}

\begin{figure}[ht]
\centering
\includegraphics[scale=0.55]{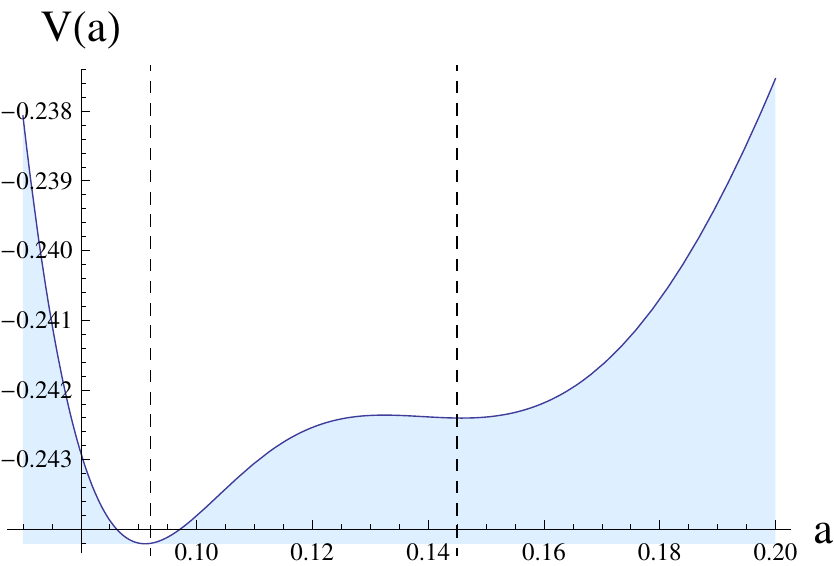}
\caption{The zoomed region of figure \ref{fig6}. The extrema of $V(a)$ are presented. The black dashed lines represent the freeze singularities.}
\label{fig11}
\end{figure}

For the case when the positive part of $f(\hat{R})$ dominates after the domination of the negative part of $f(\hat{R})$, it is possible that two freeze singularities appear. This situation is presented in figure \ref{fig4} for $f(\hat{R})=\hat{R}+10^{-2} \hat{R}^2-10^{-6}\hat{R}^3 $. In this case they appear two freeze singularities and one sudden singularity. The evolution of $b(a)+\frac{d(a)}{2}$, which corresponds with figure \ref{fig4}, is presented in figure \ref{fig5}. Note that, for values of scale factor, for which function $b(a)+\frac{d(a)}{2}$ has roots, the freeze singularities appear. $V(a)$ potential, which corresponds with figure \ref{fig4}, is presented in figure \ref{fig6} and \ref{fig11}.

\section{Singularities in the Palatini $f(R)=R+\gamma R^2+\delta R^3$ model}

For the special case of polynomial $f(\hat{R})=\hat{R}+\gamma \hat{R}^2+\delta \hat{R}^3$, one gets the following structural equation
\begin{equation}
\Omega_R-\Omega_{\delta} \Omega_R^{3}=\Omega_\text{m}+4\Omega_\Lambda,\label{structural3}
\end{equation}
where $\Omega_{\gamma} =3\gamma H_0^2$ and $\Omega_{\delta} =9\delta H_0^4$.

The Friedmann equation takes the form
\begin{multline}
\frac{H^2}{H_0^2}=\frac{b^2}{\left(b+\frac{d}{2}\right)^2}\\
\times \bigg[\frac{\Omega_\text{R}}{2b}\left[\Omega_\gamma (\Omega_\text{R}-4\Omega_\text{tot})+2\Omega_\delta \Omega_{\text{R}}(\Omega_\text{R}-3\Omega_\text{tot}))\right]\\
+\Omega_{\text{tot}}+\Omega_k\bigg],
\end{multline}
where
\begin{multline}
 \Omega_{\text{tot}} =\Omega_{m,0}a^{-3}+\Omega_{\Lambda,0},\\
b =f'(\hat{R})=1+\Omega_{\text{R}}\left[2\Omega_{\gamma}+3\Omega_{\delta}\Omega_{\text{R}}\right],\;\; \\ d=\frac{1}{H}\frac{db}{dt}=6\frac{\Omega_{\gamma}+3 \Omega_{\delta}\Omega_{\text{R}} }{3\Omega_{\delta}\Omega_{\text{R}}^2-1}\left[\Omega_{\text{R}}(1-\Omega_{\delta}\Omega_{\text{R}}^2)-4\Omega_{\Lambda,0}\right].
\end{multline}

The condition for appearance of the freeze singuarity is $b+\frac{d}{2}=0$ and in  this case it has the form
\begin{equation}
3\Omega_\gamma  \Omega_\delta  \Omega_R^3+9\Omega_\delta  \Omega_R^2+(\Omega_\gamma -36 \Omega_\delta  \Omega_\Lambda )\Omega_R-12 \Omega_\gamma  \Omega_\Lambda -1=0 \label{condition2}
\end{equation}
Equation (\ref{condition2}) has the following solution
\begin{multline}
\Omega_{\text{R}_\text{sing}}= \Omega_\gamma^{-1} \bigg[-1+\frac{r(\Omega_\gamma,\Omega_\delta,\Omega_\Lambda )}{9 2^{1/3}  \Omega_\delta }\\
-\frac{2^{1/3} \left(-81 \Omega_\delta ^2+9 \Omega_\gamma  \Omega_\delta 
(\Omega_\gamma -36 \Omega_\delta  \Omega_\Lambda )\right)}{9 r(\Omega_\gamma,\Omega_\delta,\Omega_\Lambda )  \Omega_\delta }\bigg],
\end{multline}
where
\begin{multline}
r(\Omega_\gamma,\Omega_\delta,\Omega_\Lambda )=\\ 2\Big[243 \Omega_\gamma ^2 \Omega_\delta ^2 (1+6 \Omega_\gamma  \Omega_\Lambda )-729 \Omega_\delta ^3 (1+6 \Omega_\gamma  \Omega_\Lambda )\\ 
+\Big(59049 \left(\Omega_\gamma ^2-3 \Omega_\delta \right)^2
\Omega_\delta ^4 (1+6 \Omega_\gamma  \Omega_\Lambda )^2\\
-\left(81 \Omega_\delta ^2-9 \Omega_\gamma  \Omega_\delta  (\Omega_\gamma -36 \Omega_\delta  \Omega_\Lambda )\right)^3\Big)^{1/2}\Big]^{1/3}.
\end{multline}

For the sudden singularity the condition $b=0$ provides the equation
\begin{equation} \label{eq:23}
1+\Omega_{\text{R}}\left[2\Omega_{\gamma}+3\Omega_{\delta}\Omega_{\text{R}}\right]=0.
\end{equation}
which has the following solutions
\begin{equation}
\Omega_{\text{R}_\text{sing}}=\frac{-\Omega_\gamma \pm \sqrt{\Omega_\gamma^2-3\Omega_\delta}}{3\Omega_\delta}.
\end{equation}

\section{The Palatini approach in the Einstein frame}

If $f''(\hat R) \neq 0 $ then the action (\ref{action}) can be rewritten in dynamically equivalent form of the first order Palatini gravitational action \cite{DeFelice:2010aj, Sotiriou:2008rp, Capozziello:2015wsa}
\begin{multline}\label{action1}
 S(g_{\mu\nu}, \Gamma^\lambda_{\rho\sigma}, \chi)=\frac{1}{2}\int\mathrm{d}^4x\sqrt{-g}\left(f^\prime(\chi)(\hat R-\chi) + f(\chi) \right) \\ + S_m(g_{\mu\nu},\psi),
\end{multline}
Let $\Phi=f'(\chi)$ be a scalar field, where $\chi=\hat R$. Then the action (\ref{action1}) takes the form
\begin{multline}\label{actionP}
 S(g_{\mu\nu}, \Gamma^\lambda_{\rho\sigma},\Phi)=\frac{1}{2}\int\mathrm{d}^4x\sqrt{-g}\left(\Phi \hat R - U(\Phi) \right) \\ + S_m(g_{\mu\nu},\psi),
\end{multline}
where the potential $U(\Phi)$ is given as
\begin{equation}\label{PotentialP}
 U_f(\Phi)\equiv U(\Phi)=\chi(\Phi)\Phi-f(\chi(\Phi))
\end{equation}
with $\Phi = \frac{d f(\chi)}{d\chi}$ and $\hat R\equiv \chi = \frac{d U(\Phi)}{d\Phi}$.

After the Palatini variation of the action (\ref{actionP}) we get the following equations of motion
\begin{subequations}	
 \begin{align}
	\label{EOM_P}
	\Phi\left( \hat R_{\mu\nu} - \frac{1}{2} g_{\mu\nu} \hat R \right) &  +{1\over 2} g_{\mu\nu} U(\Phi) - T_{\mu\nu} = 0,\\
	\label{EOM_connectP}
	& \hat{\nabla}_\lambda(\sqrt{-g}\Phi g^{\mu\nu})=0,\\
	%
	\label{EOM_scalar_field_P}
	  \hat R &  -  U^\prime(\Phi) =0.
	\end{align}
\end{subequations}
As a consequence of (\ref{EOM_connectP}) the connection $\hat \Gamma$ is a metric connection for a new (conformally related) metric $\bar g_{\mu\nu}=\Phi g_{\mu\nu}$; thus $\hat R_{\mu\nu}=\bar R_{\mu\nu}, \bar R= \bar g^{\mu\nu}\bar R_{\mu\nu}=\Phi^{-1} \hat R$ and $\bar g_{\mu\nu}\bar R=\ g_{\mu\nu}\hat R$.
The $g$-trace of (\ref{EOM_P}) gives a new structural equation
\begin{equation}\label{struc2}
  2U(\Phi)-U'(\Phi)\Phi=T.
\end{equation}
Equations (\ref{EOM_P}) and (\ref{EOM_scalar_field_P}) can be rewritten in the following form
	\begin{align}
	\label{EOM_P1}
	 \bar R_{\mu\nu} - \frac{1}{2} \bar g_{\mu\nu} \bar R  &  = \bar T_{\mu\nu}-{1\over 2} \bar g_{\mu\nu} \bar U(\Phi),\\
 	%
	\label{EOM_scalar_field_P1}
	  \Phi\bar R &  -  (\Phi^2\,\bar U(\Phi))^\prime =0,
	\end{align}
where $\bar U(\phi)=U(\phi)/\Phi^2$, $\bar T_{\mu\nu}=\Phi^{-1}T_{\mu\nu}$. In this case, the structural equation is given by the following formula
\begin{equation}\label{EOM_P1c}
 \Phi\,\bar U^\prime(\Phi)  + \bar T = 0\,.
\end{equation}
The action for the metric $\bar g_{\mu\nu }$ and the scalar field $\Phi$ can be recast to the Einstein frame form
\begin{equation}\label{action2}
 S(\bar g_{\mu\nu},\Phi)=\frac{1}{2}\int\mathrm{d}^4x\sqrt{-\bar g}\left(\bar R- \bar U(\Phi) \right) + S_m(\Phi^{-1}\bar g_{\mu\nu},\psi)
\end{equation}
with non-minimal coupling between $\Phi$ and $\bar g_{\mu\nu}$
\begin{equation}\label{em_2}
    \bar T^{\mu\nu} =
-\frac{2}{\sqrt{-\bar g}} \frac{\delta}{\delta \bar g_{\mu\nu}}S_m  = (\bar\rho+\bar p)\bar u^{\mu}\bar u^{\nu}+ \bar p\bar g^{\mu\nu}=\Phi^{-3}T^{\mu\nu}~,
\end{equation}
$\bar u^\mu=\Phi^{-{1\over 2}}u^\mu$, $\bar\rho=\Phi^{-2}\rho,\ \bar p=\Phi^{-2}p$, $\bar T_{\mu\nu}= \Phi^{-1}T_{\mu\nu}, \ \bar T= \Phi^{-2} T$ (see e.g. \cite{Capozziello:2015wsa, Dabrowski:2008kx}).

The metric $\bar g_{\mu\nu}$ takes the standard FRW form
\begin{equation}\label{frwb}
d\bar s^2=-d\bar t^2+\bar a^2(\bar t)\left[dr^2+r^2(d\theta^2+\sin^2\theta d\phi^2)\right],
\end{equation}
where $d\bar t=\Phi(t)^{\frac{1}{2}}\, dt$ and a new scale factor $\bar a(\bar t)=\Phi(\bar t)^{\frac{1}{2}}a(\bar t) $. 	
In the case of the barotropic matter, the cosmological equations are
\begin{equation}\label{frwb2}
3\bar H^2= \bar \rho_\Phi + \bar\rho_\text{m}, \quad 6\frac{\ddot{\bar a}}{\bar a}=2\bar\rho_\Phi -\bar{\rho}_\text{m}(1+3w)
\end{equation}
where
\begin{equation}\label{frwb3}
\bar\rho_\Phi={\frac{1}{2}}\bar U(\Phi),\quad \bar{\rho}_{\text{m}}=\rho_0\bar a^{-3(1+w)}\Phi^{\frac{1}{2}(3w-1)}
\end{equation}
and
$w=\bar p_{\text{m}} / \bar\rho_{\text{m}}= p_{\text{m}} / \rho_{\text{m}}$. In this case, the conservation equations has the following form
\begin{equation}\label{frwb4}
\dot{\bar{\rho}}_{\text{m}}+3\bar H\bar{\rho}_{\text{m}}(1+w)=-\dot{\bar{\rho}}_\Phi.\end{equation}

\begin{figure}[ht]
\centering
\includegraphics[scale=0.55]{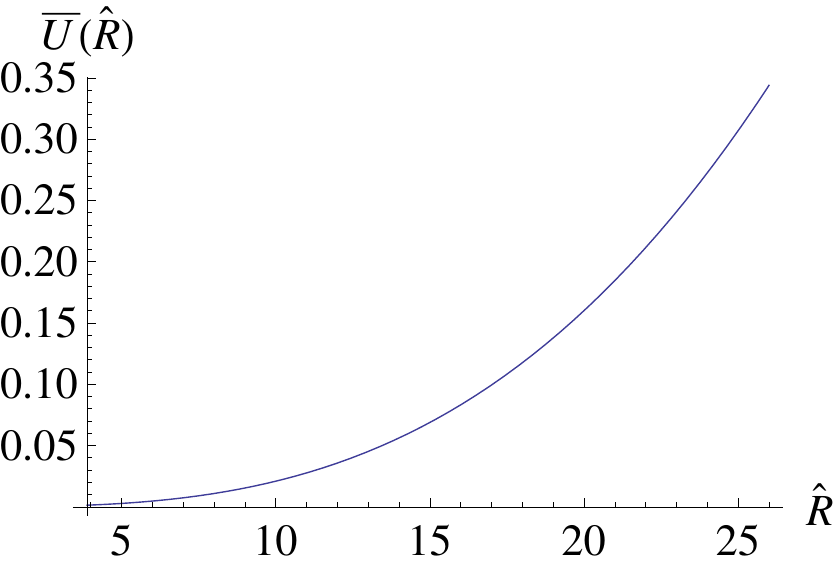}
\caption{The evolution of $\bar U(\hat R)$ in the Einstein frame in the case when the evolution of the universe starts from the generalized sudden singularity. For the illustration it is assumed that $f(\hat{R})=\hat{R}+\gamma \hat{R}^2+\delta\hat{R}^3 $, where $\gamma=10^{-9}\left[\frac{\text{s}^2\text{Mpc}^2}{\text{km}^2}\right]$ and $\delta=10^{-13}\left[\frac{\text{s}^4\text{Mpc}^4}{\text{km}^4}\right]$. Units of $\bar U(\hat R)$ are expressed in $\frac{10^4 \text{km}^2}{\text{s}^2\text{Mpc}^2}$.}
\label{fig20}
\end{figure}

Let us consider our Palatini $f(\hat{R})=\sum_{i=1}^n \gamma_i \hat{R}^i$ model in the Einstein frame, where $\gamma_1=1$. The potential $\bar U$ is given by the following formula
\begin{equation}
\bar U(\hat R)=2\bar\rho_\Phi(\hat R)=\frac{\sum_{i=1}^n (i-1)\gamma_i \hat{R}^i}{\left(\sum_{i=1}^n i\gamma_i \hat{R}^{i-1}\right)^2}.
\end{equation}
The scalar field $\Phi$ can be parametrized by $\hat R$ in the following way
\begin{equation}
\Phi(\hat R)=\frac{df(\hat R)}{d\hat R}=\sum_{i=1}^n i\gamma_i \hat{R}^{i-1}.
\end{equation}
The relation between $\bar U$ and $\hat R$ for $f(\hat{R})=\hat{R}+\gamma \hat{R}^2+\delta\hat{R}^3 $ case is presented in Fig.~\ref{fig20}.

\begin{figure}[ht]
\centering
\includegraphics[scale=0.55]{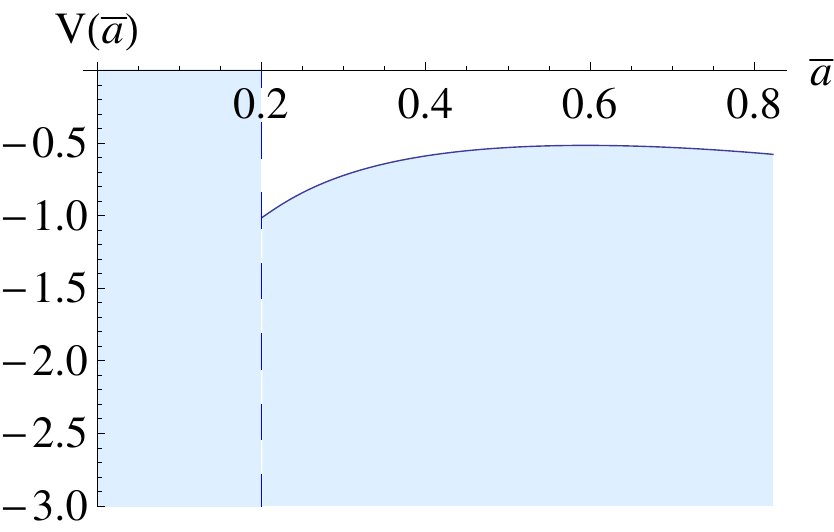}
\caption{The evolution of $V(\bar a)$ in the Einstein frame in the case when the evolution of the universe starts from the generalized sudden singularity. For the illustration it is assumed that $f(\hat{R})=\hat{R}+\gamma \hat{R}^2+\delta\hat{R}^3 $. This figure corresponds to figure \ref{fig12}. The blue dashed line presents the generalized sudden singularity. Note that the undesirable freeze singularity disappears.}
\label{fig14}
\end{figure}

\begin{figure}[ht]
\centering
\includegraphics[scale=0.55]{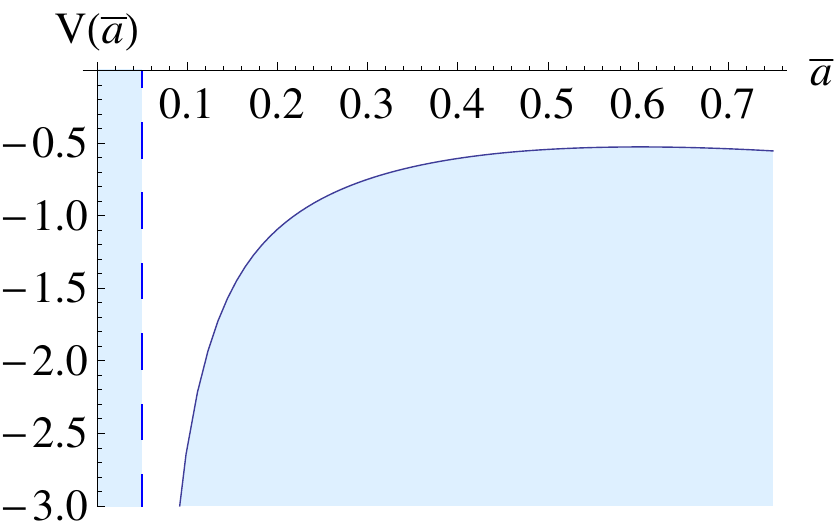}
\caption{The evolution of $V(\bar a)$ in the Einstein frame in the case when the evolution of the universe starts from the freeze singularity. For the illustration it is assumed that $f(\hat{R})=\hat{R}+\gamma \hat{R}^2+\delta\hat{R}^3 $. This figure corresponds with figure \ref{fig13}. The blue dashed line presents the freeze singularity.}
\label{fig15}
\end{figure}

\begin{figure}[ht]
\centering
\includegraphics[scale=0.5]{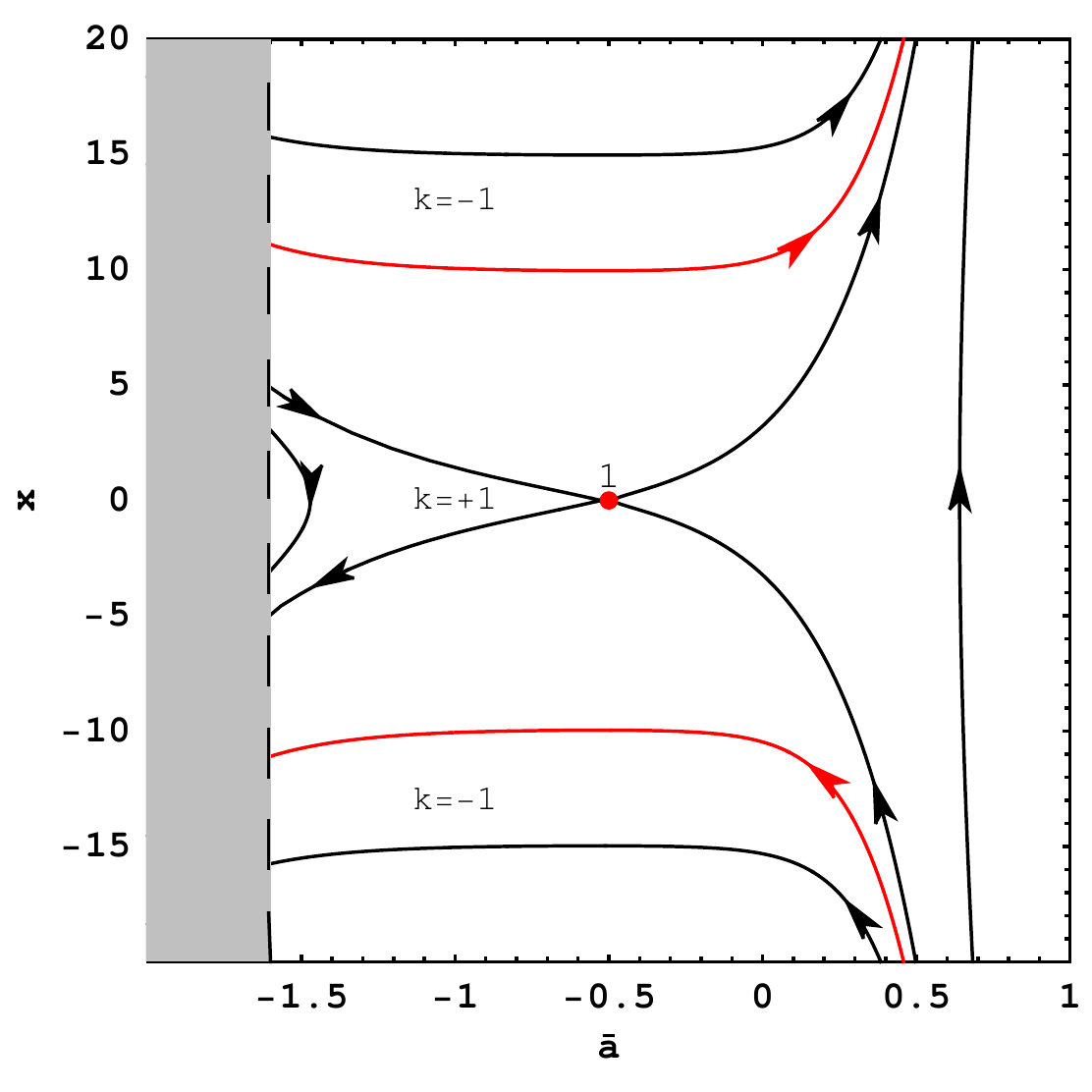}
\caption{The phase portrait for system (\ref{dota})-(\ref{dotx}) in the Einstein frame in the case when the evolution of the universe starts from the generalized sudden singularity. For the illustration it is assumed that $f(\hat{R})=\hat{R}+\gamma \hat{R}^2+\delta\hat{R}^3 $, where $\gamma=10^{-9}\left[\frac{\text{s}^2\text{Mpc}^2}{\text{km}^2}\right]$ and $\delta=10^{-13}\left[\frac{\text{s}^4\text{Mpc}^4}{\text{km}^4}\right]$. Critical point (1) represents the static Einstein universe and is a saddle. The black dashed line presents the generalized sudden singularity. The grey color presents the non-physical domain ($\bar a<\bar a_\text{s}$). The red trajectories represent the path of evolution for the flat universe. These trajectories seperate the domain with the negative curvature ($k=-1$) from the domain with the positive curvature ($k=+1$). The scale factor is expressed in the logarithmic scale.}
\label{fig12}
\end{figure}

\begin{figure}[ht]
\centering
\includegraphics[scale=0.5]{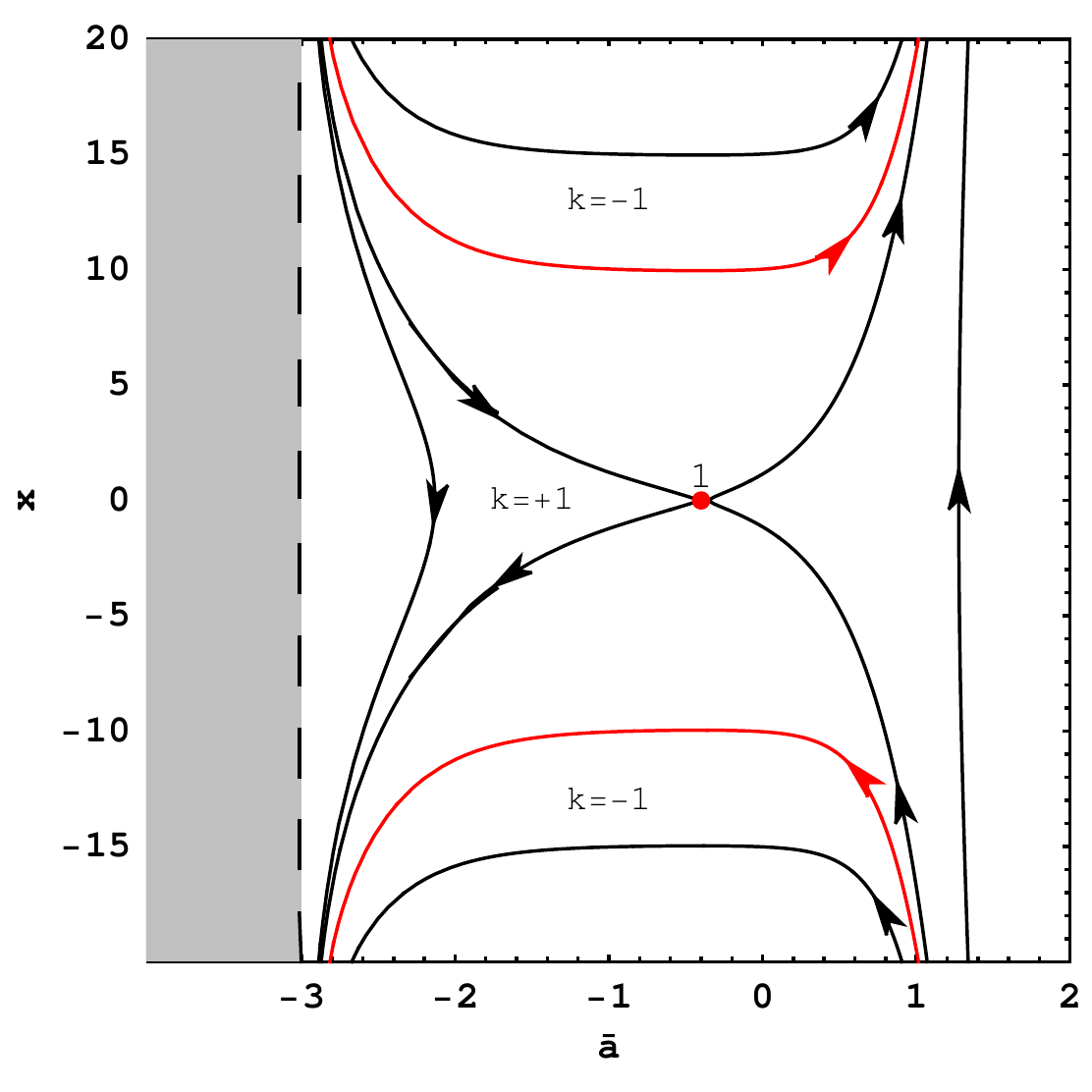}
\caption{The phase portrait for system (\ref{dota})-(\ref{dotx}) in the Einstein frame in the case when the evolution of the universe starts from the freeze singularity. For the illustration it is assumed that $f(\hat{R})=\hat{R}+\gamma \hat{R}^2+\delta\hat{R}^3 $, where $\gamma=-10^{-9}\left[\frac{\text{s}^2\text{Mpc}^2}{\text{km}^2}\right]$ and $\delta=-10^{-13}\left[\frac{\text{s}^4\text{Mpc}^4}{\text{km}^4}\right]$. Critical point (1) represents the static Einstein universe and is a saddle. The black dashed line presents the freeze singularity. The grey color presents the non-physical domain ($\bar a<\bar a_\text{s}$). The red trajectories represent the path of evolution for the flat universe. These trajectories separate the domain with the negative curvature ($k=-1$) from the domain with the positive curvature ($k=+1$). The scale factor is expressed in the logarithmic scale.}
\label{fig13}
\end{figure}

In this frame, two scenarios of cosmic evolution may appear. In the first one the evolution of the universe starts from the generalized sudden singularity. The second case is when it starts from the freeze singularity. The diagrams of the corresponding Newtonian potentials $V(\bar a)$ are presented in Figs.~\ref{fig14} and \ref{fig15}. We can use the potential $V(\bar a)$ to construct phase space portraits analogous to the ones in section 3 (see Figs.~\ref{fig12} and \ref{fig13}).

\begin{figure}[ht]
\centering
\includegraphics[scale=0.55]{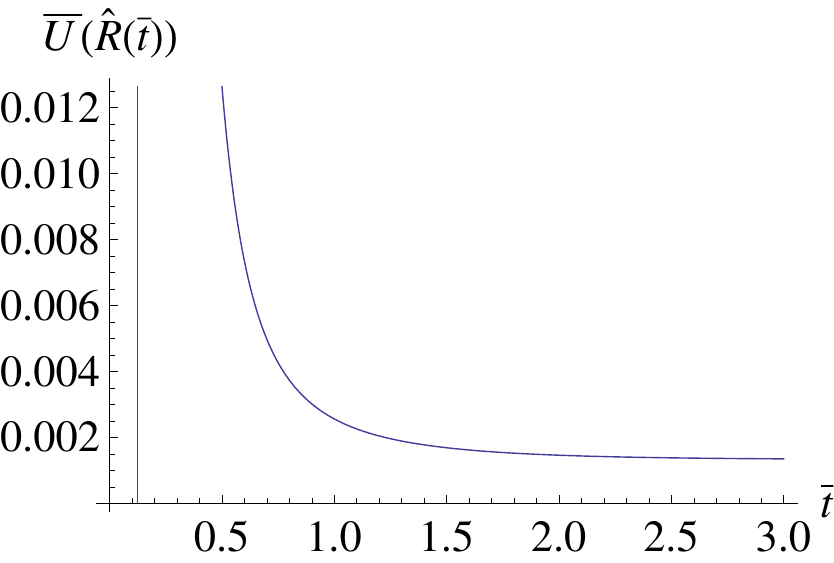}
\caption{The evolution of $\bar U(\hat R(\bar t))$ in the Einstein frame in the case when the evolution of the universe starts from the generalized sudden singularity. For the illustration it is assumed that $f(\hat{R})=\hat{R}+\gamma \hat{R}^2+\delta\hat{R}^3 $, where $\gamma=10^{-9}\left[\frac{\text{s}^2\text{Mpc}^2}{\text{km}^2}\right]$ and $\delta=10^{-13}\left[\frac{\text{s}^4\text{Mpc}^4}{\text{km}^4}\right]$. Note that for the late time $\bar U(\bar t)$ potential goes to a constant value at late time. Units of time are expressed in $\frac{\text{s Mpc}}{\text{100 km}}$ and units of $\bar U(\hat R(\bar t))$ are expressed in $\frac{10^4 \text{km}^2}{\text{s}^2\text{Mpc}^2}$.}
\label{fig18}
\end{figure}

\begin{figure}[ht]
\centering
\includegraphics[scale=0.55]{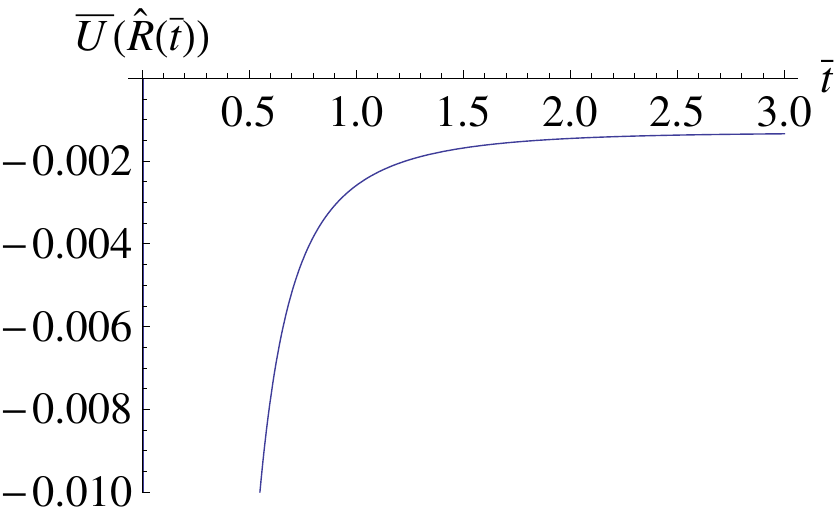}
\caption{The evolution of $\bar U(\hat R(\bar t))$ in the Einstein frame in the case when the evolution of the universe starts from the freeze singularity. For the illustration it is assumed that $f(\hat{R})=\hat{R}+\gamma \hat{R}^2+\delta\hat{R}^3 $, where $\gamma=-10^{-9}\left[\frac{\text{s}^2\text{Mpc}^2}{\text{km}^2}\right]$ and $\delta=-10^{-13}\left[\frac{\text{s}^4\text{Mpc}^4}{\text{km}^4}\right]$. Units of time are expressed in $\frac{\text{s Mpc}}{\text{100 km}}$ and units of $\bar U(\hat R(\bar t))$ are expressed in $\frac{10^4 \text{km}^2}{\text{s}^2\text{Mpc}^2}$.}
\label{fig17}
\end{figure}

The evolution of the scalar field potential $\bar U(\bar t)$, which plays a role of dynamical cosmological constant, is presented in Fig.~\ref{fig18} for the case with the generalized sudden singularity. Note that for the late time the potential $\bar U(\bar t)$ is constant. The evolution of $\bar U(\bar t)$, for the case when the freeze singularity appears, is presented in Fig.~\ref{fig17}. For the late time the potential $\bar U(\bar a)$ can be approximated as
\begin{equation}
\bar U(\bar a)=\frac{2\gamma \hat R(\bar a)^2}{1+6\gamma \hat R(\bar a)}=\frac{2\gamma(4\Lambda+\bar\rho_{\text{m},0}\bar a^{-3})}{1+6\gamma(4\Lambda+\bar\rho_{\text{m},0}\bar a^{-3})}.
\end{equation}

\begin{figure}[ht]
\centering
\includegraphics[scale=0.55]{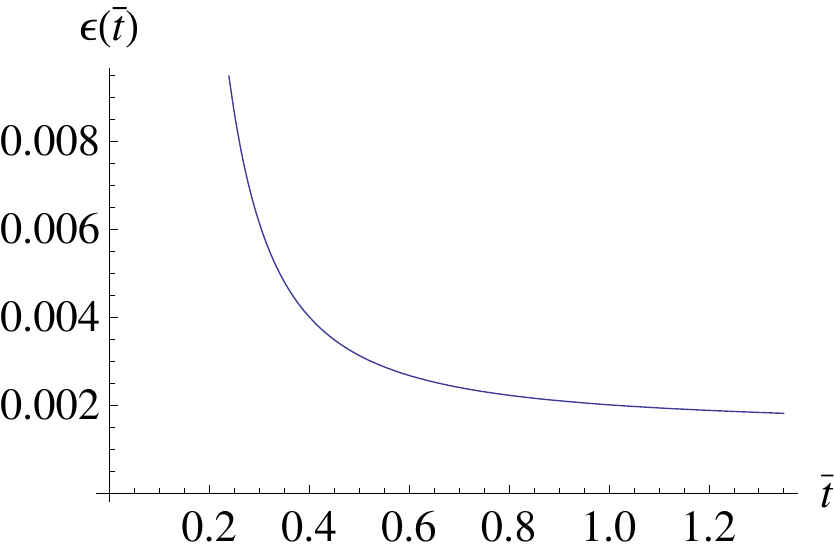}
\caption{The evolution of $\epsilon(\bar t)$ in the Einstein frame in the case when the evolution of the universe starts from the generalized sudden singularity. For the illustration it is assumed that $f(\hat{R})=\hat{R}+\gamma \hat{R}^2+\delta\hat{R}^3 $, where $\gamma=10^{-9}\left[\frac{\text{s}^2\text{Mpc}^2}{\text{km}^2}\right]$ and $\delta=10^{-13}\left[\frac{\text{s}^4\text{Mpc}^4}{\text{km}^4}\right]$. Note that for the late time $\epsilon(\bar t)$ is constant. Units of time are expressed in $\frac{\text{s Mpc}}{\text{100 km}}$.}
\label{fig19}
\end{figure}

From the structural equation (\ref{EOM_P1c}) for $f(\hat{R})=\hat{R}+\gamma \hat{R}^2+\delta\hat{R}^3 $ case, we get the parameterization of the dust matter density with respect to $\hat R$
\begin{equation}
\bar\rho_{\text{m}}=\frac{\hat R-\delta \hat R^3}{(1+2\gamma \hat R+3\delta \hat R^2)^2}-4\Lambda.
\end{equation}
It is interesting that in the Einstein frame the interaction between dark matter and dark energy naturally appears as a physical phenomenon. This interaction modifies the original scaling law for dust matter by a function $\epsilon(\bar t)$.  
\begin{equation}
\bar\rho_{\text{m}}=\bar\rho_{\text{m},0} \bar a(\bar t)^{-3+\epsilon(\bar t)},
\end{equation}
where $\epsilon=\frac{1}{\ln \bar a}\int{\frac{Q}{\bar H\bar\rho_\text{m}}}d\ln \bar a$ and $Q=-\dot{\bar\rho}_{\phi}=\bar H(\hat R)\bar\rho_\text{m}(\hat R)\frac{3\hat{R}(\gamma+3\delta\hat{R})(\delta\hat{R}^2-1))}{\hat{R}(\gamma+3\delta \hat{R}(3+\gamma \hat{R}))-1}$ for $f(\hat{R})=\hat{R}+\gamma \hat{R}^2+\delta\hat{R}^3 $ case.
The evolution of $\epsilon(\bar t)$ is presented in Fig.~\ref{fig19}.

\section{Conclusions}

The main goal of the paper was to point out some advantages of formulation the Palatini FRW cosmology in the Einstein frame. The most crucial one is that in the Einstein frame the parametrization of dark energy is uniquely determined. In general it is obtained in the covariant form as a function of the Ricci scalar.

It is well know that scalar-tensor theories of gravity can be formulated both in the Jordan as well as in the Einstein frame. These frames are conformally related \cite{Weinberg:1972}.
We also know that the formulations of a scalar-tensor theory in  two different conformal frames although mathematically equivalent are physically inequivalent.

Faraoni and Gunzing gives a simple argument which favours the Einstein frame over the Jordan frame because in the latter one should potentially detect the time–dependent amplification induced by gravitational waves \cite{Faraoni:1999hp}.

An analogous problem has been dedected in $f(R)$ gravity that the Jordan frames could be physically non-equivalent, although they are connected by a conformal transformation \cite{Capozziello:2010sc, Bahamonde:2016wmz}.
In principle, there are two types of admissible arguments for favouring one frame over another one: the one  coming from observations (for example astronomical observations) or theoretical nature (e.g. showing that some obstacles or pathologies will vanishing in privileged frame).

From our investigation of the model in Einstein frame we obtained that some pathologies like degenerated multiple freeze singularities \cite{Stachowski:2016zio} disappear in a generic case. The big bang singularity is replaced by the singularity of finite scale factor. Because the potential $\bar U(\hat R(\bar t))$ is constant for the late time,   in the case when matter is negligible, the inflation appears like in the case $f(\hat R)=\hat R+\gamma \hat R^2$ \cite{Stachowski:2016zio}.

There are also some other advantages when transforming to Einstein frame, namely that in this frame one naturally obtains the formula on dynamical dark energy which  is going at late time toward cosmological constant. It is important that corresponding parametrization of dark energy is not postulated ad hock but it emerges from the first principles -- which is the formulation of the problem in the Einstein frame. It is important that  the parametrization of dark energy (energy density as well as a pressure) in terms of the Ricci scalar is given in a covariant form from the structure equation.

After transition to the Einstein frame the model evolution is governed by the Friedmann equation with two interacting fluids: dark energy and dark matter. 
This interaction modifies the standard scaling of the redshift relation for dark matter.

\providecommand{\href}[2]{#2}\begingroup\raggedright\endgroup

\end{document}